# Multibeam Phased Arrays with Spherical-Gold Spatio-temporal Coding for Fading-Resilient and Delay-Robust Beam Isolations

Yuan Ma, *Graduate Student Member*, *IEEE*, Mike Ballou, Kyle Richard, Hessam Mahdavifar, *Member, IEEE*
Najme Ebrahimi, *Member, IEEE*

*Abstract*— Future integrated sensing and communication (ISAC) systems require simultaneous multibeam operation with low-latency hardware and robust isolation under synchronization error and fading. Conventional code-division multiplexing using Walsh–Hadamard codes is extremely time-sensitive. This paper demonstrates that conventional temporal-only coded multibeam arrays, results in inter-beam sidelobe level (SLL) collapsing to within a few dB of the main lobe and varying by more than 10–20 dB over delay. By embedding moderate-length Gold sequences into a spherical spatial codebook, the proposed Spherical-Gold scheme leverages both $1/\sqrt{N}$ temporal and $1/\sqrt{M}$ spatial bounds, achieving effectively $1/(\sqrt{N}\cdot\sqrt{M})$ correlation without increasing RF complexity. The measurement results and verifications have been performed using Analog Device, ADAR3002, Ka-band 256-element receiver with four simultaneous beams demonstrate ≥15 dB rejection with <±2.5 dB variation of SLL under time-error and fading, while temporal-only CDMA degrades toward (-5 to -7 dB SLL and near <±8 dB variations under time-delay.

## I. Introduction

Next-generation integrated sensing and communication systems require multibeam and multiplexing for low latency, high throughput, and low-complexity scalable antenna array [1-3]. For example, one beam can be dedicated to data communication while another beam is simultaneously used for sensing and angle/direction of arrival (AoA/DoA) detection, situational awareness, Fig. 1, [1-3], or physical layer security, [4-6]. Previous works have used code-multiplexing for simultaneous multibeam phased-array receivers, [7], multi-channel systems with a shared IF chain and shared ADC [8-10], as well as full-duplex system [8, 9]. Each antenna element or subarray is encoded with a code-division multiple-access (CDMA) sequence such as Walsh-Hadamard to separate channels and provide isolation between elements. However, the main challenge for such CDMA-based systems is time-delay sensitivity. Small synchronization errors and fading destroy orthogonality and can reduce isolation dramatically [11, 12]. For Walsh-Hadamard codes, a one-chip timing error in an 8-chip sequence can raise the normalized cross-correlation between codes to ~0.8-0.9, which corresponds to only 1-2 dB rejection (the "false" peak is only 1 dB below the intended main peak), Fig. 2. This means that a strong interferer encoded with another Walsh code can leak almost fully into the desired beam when there is a modest delay mismatch. We define this leakage as inter-beam side-lobe level (SLL). Therefore, cross correlation $\rho_{ij}^{code}(\tau)$ between the codes $C_i(t)$ and $C_j(t)$ is used to estimate inter-beam isolation, while the autocorrelation $\rho_{ii}^{code}(\tau)$ is used for matched filtering and sensing (detection of the intended coded beam, AoA sensing, etc.), Fig. 1.

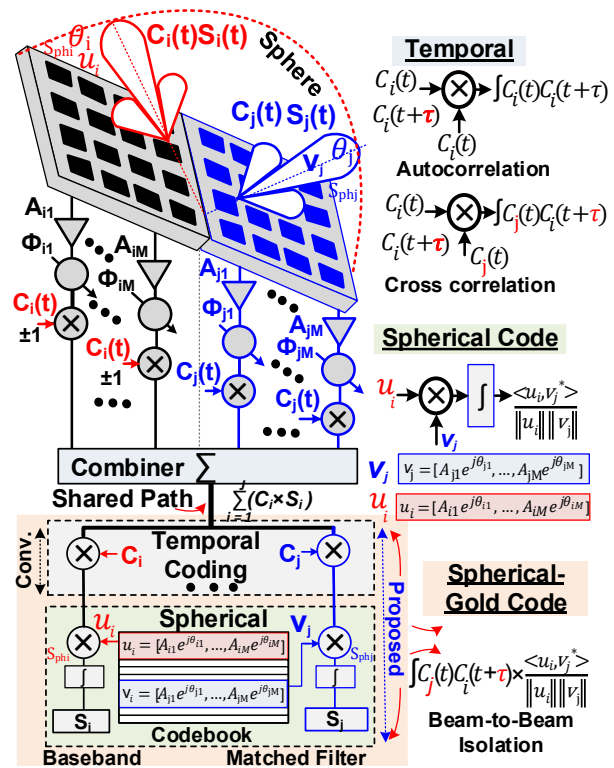

**Fig.1**: Conceptual Spherical-Gold multibeam phased-array architecture for ISAC. Each subarray beam, $S_i(t)$, is temporally encoded with a code $C_i(t)$, and decoded in baseband using autocorrelation for sensing and cross-correlation for beam separation. The proposed Spherical-Gold processing combines temporal coding with the spherical codewords to achieve delay and fading-resilient multibeam isolation.

This paper presents a new paradigm for encoding and decoding multibeam arrays using both spatial (spherical) and temporal (Gold) coding to tackle this long-standing fading and timing-error problem. The proposed Spherical-Gold code makes the system fading-resilient by combining Gold codes, which maintain low and bounded cross-correlation under timing misalignment (typical worst-case ≈ –7 to –10 dB instead of –1 dB for Walsh), and a spatial spherical codebook constructed from the array beamforming weights and geometry, which adds an extra spatial filter for beam-to-beam and channel-to-channel isolation (>15 dB) under time delay.

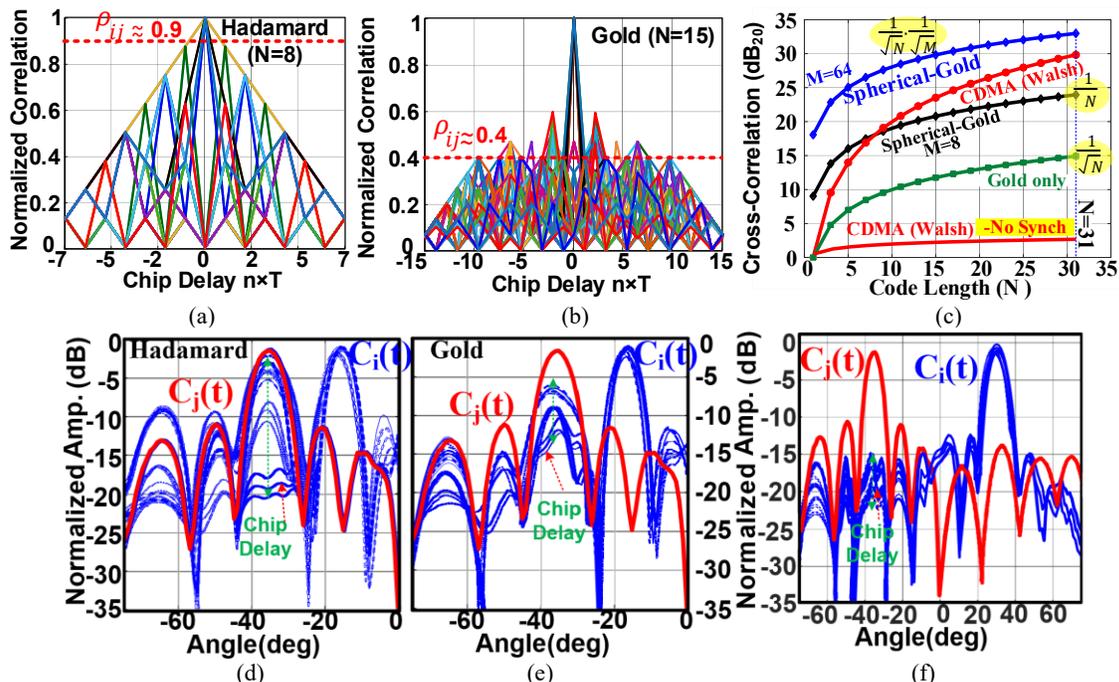

Fig. 2: Simulation and theory analysis for Temporal code and Spherical-Gold code and isolation bounds: (a) correlation of Hadamard and (b) correlation of Gold codes under fading-induced chip delays, (c) theoretical cross-correlation and maximum bound for inter-beam isolation between Gold $1/\sqrt{N}$, Hadamard, $1/N$ and proposed Spherical-Gold, $1/(\sqrt{N}\cdot\sqrt{M})$, highlighting superior performance of the proposed Spherical-Gold code. Simulated inter-beam sidelobe levels for two 64-element nested subarrays under chip-level timing error. (d) Hadamard coding, (e) Gold code, and (f) proposed Spherical-Gold code with two beam set [$\theta(u_i)$ =-35°, $\theta(v_i)$ =35°], showing SLL<-15 dB.

Sec. II presents the theoretical analysis and simulation results comparing conventional temporal multiplexing schemes (Walsh/Hadamard and Gold) with the proposed Spherical-Gold coding. Section III provides measured results demonstrating the performance of the Spherical-Gold scheme using a Keysight Compact Antenna Test Range (CATR) setup with a 256-element Ka-band phased array receiver based on Analog Devices' ADAR3002. Sec. IV concludes the paper.

## II. PROPOSED TIME-ERROR AND FADING-RESILIENT MULTIBEAM MULTIPLEXING (SPHERICAL-GOLD CODING)

### A. Conventional temporal coding

For a scalable array for multiplexing of $J$ multibeam and total array elements $M$, each subarray consists of $m = M/J$ elements. The antenna selection can be done using Sparse or thinned array selection to create narrow beam with spatial and geometric diversity, [7]. In conventional multiplexing techniques, each subarray is first encoded with a temporal code $C_i(t)$, where $i = 1, ..., J$, and the code length is $N$ chips. At the receiver, baseband (BB) processing performs matched filtering by correlating the received signal with the intended code, Fig. 1. The autocorrelation $\rho_{ii}(\tau)$ is used for processing the desired beam, while the cross-correlation $\rho_{ij}(\tau)$ is used to evaluate inter-beam isolation and unintended beam leakage, as illustrated in Fig. 1 and the equation below:

$$\rho_{ij}(\tau) = \int C_i(t)C_j(t+\tau)dt \approx \frac{1}{\sqrt{N}} \,(Gold); \frac{1}{N} \,(Walsh) \quad (1)$$

As shown in (1) and Fig. 2(a) and (b), ideal Hadamard codes are orthogonal at $\tau = 0$, with $\rho_{ij}(0) = 0$. However, under a delay $\tau = nT_{chip}$, the cross-correlation magnitude can rise to around 0.9, corresponding to only –1 dB isolation. In contrast, Gold codes exhibit bounded cross-correlation under delay, typically around 0.5 ($\approx$ –6 dB). Gold codes are generally constructed by combining two $m$-sequences, and their key property is that the worst-case cross-correlation scales as approximately $1/\sqrt{N}$, compared to $1/N$ for Hadamard codes. Fig. 2(c) illustrates the maximum achievable isolation for various code lengths $N$. However, to achieve isolation greater than 15 dB, Gold codes require longer sequences (e.g., $N = 31$), which increases latency and codeword overhead.

### B. Proposed Spherical-Gold code

To avoid the long-code penalty and further suppress interference, we proposed to combine temporal Gold coding with a spherical spatial code formed by the array beamforming weights. For each subarray/beam $i$, we define the complex weight vector as $u_i = [A_{i1}e^{j\theta_{i1}}, ..., A_{iM}e^{j\theta_{iM}}]^T$, where $A_{im}$ and $\theta_{im}$ are the amplitude and phase applied to element $m$ for beam $J$. This is exactly the beamforming vector used to steer to angle $\theta_i$. Each such vector acts as a spherical codeword in the spatial domain (a point on the complex unit sphere, shown in Fig. 1) and is stored in a codebook [13]. The spatial isolation between two beams with codewords $\mathbf{u}_i$ and $\mathbf{v}_j$ is defined as their normalized inner product:

$$\rho_{sph}(i,j) = \frac{\langle u_i, v_j^*\rangle}{\|u_i\|\|v_j\|} = \geq \frac{1}{\sqrt{M}} \quad (2)$$

For well-designed sparse 64-element subarrays, one can typically obtain $|\rho_{ij}^{sph}|$ in the range $\approx$ 0.12–0.2, corresponding to −18 to −14 dB spatial isolation. This is consistent with spherical code theory where the best possible bound scales like $1/\sqrt{M}$ [13], [14]. When both temporal Gold coding and spatial spherical coding are used, the array response for beam $j$ after decoding with code $C_i$ can be product of time and space:

$$\rho_{ij}(\tau) = \int C_i(t) C_j(t+\tau) dt \times \frac{\langle u_i, v_j^* \rangle}{\|u_i\|\|v_j\|} \approx \frac{1}{\sqrt{N}} \cdot \frac{1}{\sqrt{M}} \quad (3)$$

The time-domain matched filter with $C_i(t)$ contributes a factor $\rho_{ij}^{code}(\tau)$, $1/\sqrt{N}$ while the spatial overlap between beams (i.e., subarray steering vectors) contributes $\rho_{ij}^{sph}$, $1/\sqrt{M}$. As shown in Fig. 2(c), for example, with $M = 64$ (sparse 2-D array) and $N = 15$ of Gold results in isolation bound of 30 dB). Fig. 2 (d)-(f) show the simulation results for two simultaneous 64-element nested subarrays each encoded with a 15-bit temporal codeword, Gold or Hadamard, of $C_i(t)$ and $C_j(t)$. Gold and Walsh codes are implemented using ±1 symbols (0° and 180° phase shifts) such as BPSK modulation, [3]. The spherical code is implemented by choosing beamforming amplitude and phase weights from the codebook, without additional RF hardware to be needed beyond the normal beam steering. The simulations show that the SLL (inter-beam leakage) for Hadamard coding varies widely with chip delay, from about –20 dB up to –1 dB relative to the main beam, Fig. 2(d), while for Gold coding it is confined to roughly –15 to –7 dB, Fig. 2(e). In contrast, Fig. 2(f) shows that the proposed Spherical-Gold code, with two beam set of [$\theta(u_i) = -35°$, $\theta(v_i) = 35°$], limits the variation to only about 5 dB over chip delay, with the SLL in a narrower range of approximately –20 to –15 dB, that can be improved by optimized sparse array.

### III. MEASUREMENT RESULTS

To verify the proposed multibeam processing technique, we employed Keysight CATR chamber and a Ka-band receive array based on Analog Devices' ADAR3002, Fig. 3(a). The receive array under test is implemented with ADAR3002 Ka-band dual-polarized beamformers (4 elements, 2 beams) tiled to form a 256-element array. The IC covers 17.7–21.2 GHz and provides 0-360° phase control in 5.625° steps together with ≥31 dB gain control at ≤0.5 dB resolution; typical RMS phase/gain errors are ~2.3°/0.2 dB, cross-polarization isolation is ~25 dB, and on-chip memory supports up to 256 prestored beam states. This enables repeatable, fine-grained Spherical codeword programing during measurement. For the temporal experiment, each beam is modulated with a 15-chip Gold sequence. For every chip of the sequence, the CATR positioner sweeps $\theta$ from −75° to +75° at a fixed $\phi$, while the array is programmed with a particular spherical codeword. The Keysight system exports one combine received waveform into $\theta \times 15$ matrix and then apply the proposed decoding technique to process each intended beam based on Fig. 1 and (1) to (3). Although the proposed technique requires an additional matched filter (for the spherical code) in BB, such matched filtering is already needed for AoA estimation and codebook-based beam management. We simply co-design the spatial codebook and temporal Gold codes so that the same processing provides both AoA information and time-error-resilient multibeam isolation. This correlation across chips separates the four spherical subarray beams from the single IF output and yields one decoded θ-cut per beam for every imposed delay. Fig. 3(b) illustrates the captured four simultaneous beam waveforms under different steering conditions. For Spherical-coding test, the array is partitioned into four sparse subarrays and nominal beam angles at –35°, –15°, +15°, and +35°. For each beam pair we measure decoded beam patterns under time delay. The measurement results compare Temporal only decoding (Fig.3 (c) and (d)) to full Spherical-Gold decoding

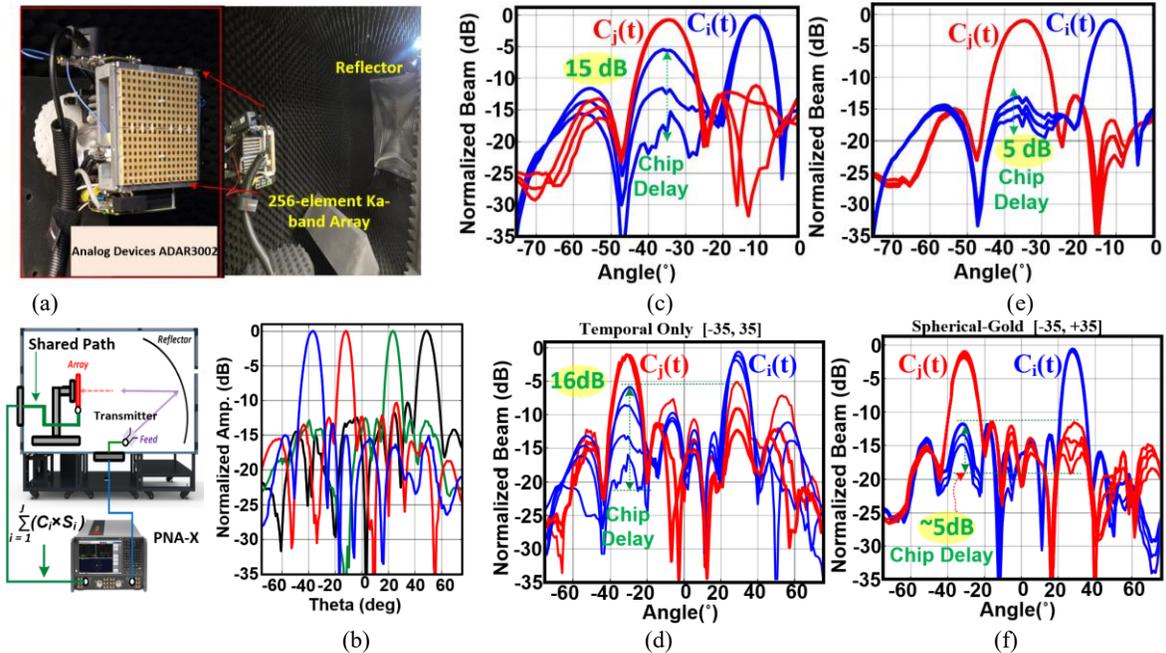

**Fig. 3:** Measurement setup using 256-element ADAR3002-based Ka-band receive array mounted in the Keysight CATR chamber, (b) Four simultaneous beam using sparse array optimization with ~16 dB SLL. (c)-(f): Measured inter-beam isolation versus chip delay for different coding and spherical angles: (c) temporal coding with sphere beam angles of [-35°, -12°], (d) temporal coding with sphere beam angles of [-35°, 35°], (e) proposed Spherical-Gold scheme with two beam angles of [-35°, -12°], (e) Spherical-Gold scheme with two angles of [-35°, 35°].

Table I: Comparison with prior multibeam and multiplexing Array

| Reference | UCSD [7] | UCSB [8] | Columbia [9] | This Work |
|---|---|---|---|---|
| #Beam/Function | 4/ Array (RX) | 1/ Full Duplex | 4/ Array (RX) | 4/ Array (RX) |
| Coding Technique | Hadamard 16-bit | Walsh 16-bit | Walsh 16-bit | Spherical-Gold N=15, M=64 |
| Isolation SLL (dB) [3] | 20-40*[1] | 15-23*[2] | 15-35*[2] | 12-35*[4] |
| Isolation w/Time Delay [3] | No/Data | No/Data | No/Data | >15 dB ±2.5 dB*[4] |
| Cross Correlation Bound | $1/N$ | $1/N$ | $1/N$ | $1/(\sqrt{N}\cdot\sqrt{M})$ |
| Fading Resilient | No | No | No | **Yes** |

*1: No fading effect, *2: Based on Simulation results *3: (Main beam – max leakage from other beam), *4: Measured results, $N$: codeword, $M$: #antenna

(Fig.3 (e) and (f)) for representative two-beam pairs. The temporal-only coding produces up to ~15–16 dB variation in isolation with delay. For only Gold coding, when two beams are close in angle, [-35, -12] the leakage sidelobe inside the other beam's main lobe can vary between about −7 dB and −16 dB across chip delay due to unoptimized Spherical codeword. Therefore, by employing spherical code and the codeword optimization, the results flatten this variation to about 5 dB and keeps the sidelobe leakage close to −15 dB across the tested delay range, Fig. 3 (e), (f). Measured isolation is limited by the sparse array design and multibeam calibration, but improved optimization can push it toward the ~30 dB theoretical bound. Table. I shows a comparison table versus prior multibeam and code-multiplexed arrays (UCSD nested Hadamard array [7], code-domain shared-IF array [9] and coding for full duplexing [8]). Prior works report 20-40 dB sidelobe isolation under ideal calibration or simulation and synchronous conditions, but do not consider isolation under time delay and fading. This work provides up to 4 simultaneous beams with better than -15 side lobe and less than <5 dB measured variation of isolation under time delay.

## IV. CONCLUSION

This paper has demonstrated that temporal-only CDMA techniques (Walsh/Hadamard or Gold) are inherently fragile to chip-level delay and fading in multibeam arrays, resulting in inter-beam sidelobe levels collapsing to within a few dB of the main lobe and varying by more than 16 dB across delay. By embedding moderate-length Gold sequences into a spherical spatial codebook, the proposed Spherical-Gold scheme leverages both $1/\sqrt{N}$ temporal and $1/\sqrt{M}$ spatial correlation bounds, achieving effective $1/(\sqrt{N}\cdot\sqrt{M})$ isolation without added RF complexity. Measurement results using a 256-element Ka-band phased array (ADAR3002-based) and Keysight CATR verification confirm that four simultaneous beams achieve sidelobes consistently below −15 dB with <5 dB variation over delay. To the best of our knowledge, this is the first work to experimentally propose and develop a novel fading- and delay-resilient multibeam coding framework that scales across multiplexed phased-array elements and supports joint communication and sensing.